\documentclass[a4paper]{scrartcl}
\pdfoutput=1
\usepackage[utf8]{inputenc}

\usepackage[T1]{fontenc} 
\usepackage{lmodern} 
\usepackage[sc,osf]{mathpazo} 
\usepackage{myT1classico} 
\DeclareMathAlphabet{\mathsf}{T1}
  {\sfdefault}{m}{n} 
\SetMathAlphabet{\mathsf}{bold}{T1}{\sfdefault}{b}{n} 

\usepackage{amsmath,amssymb,amsthm}
\usepackage{mathtools}
\usepackage{tensor}
\usepackage[Smaller]{cancel}
\usepackage{accents}

\usepackage{enumitem}

\usepackage{relsize}
\newcommand{\acronym}[1]{\texorpdfstring{\textsmaller{#1}}{#1}}

\usepackage[british]{babel}

\usepackage{microtype} 

\usepackage{authblk}

\usepackage[colorlinks]{hyperref}
\usepackage[nameinlink]{cleveref}

\DeclareFontFamily{U}{matha}{\hyphenchar\font45}
\DeclareFontShape{U}{matha}{m}{n}{
  <5> <6> <7> <8> <9> <10> gen * matha
  <10.95> matha10 <12> <14.4> <17.28> <20.74> <24.88> matha12
}{}
\DeclareSymbolFont{matha}{U}{matha}{m}{n}
\DeclareMathSymbol{\oleft}{2}{matha}{"68}
\DeclareMathSymbol{\oright}{2}{matha}{"69}

\usepackage[style=phys,biblabel=brackets,eprint=true,sorting=noney,
  sortcites=false]{biblatex}
\newbibmacro{string+doiurl}[1]{%
  \iffieldundef{doi}{%
    \iffieldundef{url}{#1}{\href{\thefield{url}}{#1}}%
  }{\href{https://doi.org/\thefield{doi}}{#1}}%
}
\DeclareFieldFormat{title}{\usebibmacro{string+doiurl}{\mkbibemph{#1}}}
\DeclareSortingTemplate{noney}{
  \sort{\citeorder}
}

\newcommand{\e}{\mathrm{e}} 
\newcommand{\D}{\mathrm{d}} 
\newcommand{\Ad}{\mathrm{Ad}}
\newcommand{\R}{\mathbb R}
\newcommand{\hv}{\underaccent{v}{h}} 
\newcommand{\Gammav}{\accentset{v}{\Gamma}}
\newcommand{\nablav}{\accentset{v}{\nabla}}
\newcommand{\Aut}{\mathsf{Aut}}
\newcommand{\GL}{\mathsf{GL}}

\newcommand{\Gal}{\mathsf{Gal}}
\newcommand{\gl}{\mathfrak{gl}}
\newcommand{\so}{\mathfrak{so}}
\newcommand{\gal}{\mathfrak{gal}}
\newcommand{\bomega}{{\boldsymbol{\omega}}}
\newcommand{\bhatomega}{{\boldsymbol{\hat\omega}}}

\theoremstyle{plain}
\newtheorem{theorem}{Theorem}
\newtheorem{lemma}[theorem]{Lemma}
\newtheorem{proposition}[theorem]{Proposition}

\theoremstyle{definition}
\newtheorem{definition}[theorem]{Definition}

\numberwithin{equation}{section}

\title{The classification of general affine connections in
  Newton--Cartan geometry}

\subtitle{Towards metric-affine Newton--Cartan gravity}

\author{Philip K. Schwartz}
\affil{Institute for Theoretical Physics,
  Leibniz University Hannover, \par
  Appelstraße 2, 30167 Hannover, Germany}
\affil{\normalfont\texttt{\href{mailto:philip.schwartz@itp.uni-hannover.de}
    {philip.schwartz@itp.uni-hannover.de}}}
 
\date{}

\begin{document}
\maketitle

\begin{abstract}
  \noindent
  We give a full classification of general affine connections on
  Galilei manifolds in terms of independently specifiable tensor
  fields.  This generalises the well-known case of (torsional) Galilei
  connections, i.e.\ connections compatible with the metric structure
  of the Galilei manifold.  Similarly to the well-known
  pseudo-Riemannian case, the additional freedom for connections that
  are not metric-compatible lies in the covariant derivatives of the
  two tensors defining the metric structure (the clock form and the
  space metric), which however are not fully independent of each
  other.
\end{abstract}

\section{Introduction}
\label{sec:intro}

It is well-known that on a pseudo-Riemannian manifold $(M,g)$ any
affine connection $\nabla$ is uniquely determined by its torsion $T$
and non-metricity $\nabla g$.  This allows for the reformulation of
standard general relativity (\acronym{GR}) \cite{MTW:2017,
  Straumann:2013}, in which gravitational and inertial effects are
encoded in the torsion-free metric-compatible (and, generally, curved)
Levi-Civita connection of a Lorentzian metric $g$, either in terms of
flat torsionful metric-compatible connections, giving rise to the
(metric) \emph{teleparallel equivalent of \acronym{GR}}
(\acronym{TEGR}) \cite{Maluf:2013}, or in terms of flat torsion-free
non-metric connections, giving rise to the \emph{symmetric
  teleparallel equivalent of \acronym{GR}} (\acronym{STEGR})
\cite{Nester.Yo:1999}.  These three equivalent formulations of
\acronym{GR}, termed the `geometric trinity of gravity'
\cite{Jimenez.EtAl:2019}, allow for different generalisations, i.e.\
modified theories of gravity.  The general framework of such
generalised geometric theories of `relativistic' gravity in which the
geometric fields are a Lorentzian metric as well as an independent,
(initially) completely arbitrary affine connection is called
\emph{metric-affine gravity}.

The Newtonian $c\to\infty$ limit of standard \acronym{GR} may be
formulated in differential-geometric terms, giving rise to
Newton--Cartan gravity \cite{Cartan:1923_24,Cartan:1986,
  Friedrichs:1928,Trautman:1963,Trautman:1965,Dombrowski.Horneffer:1964,
  Kuenzle:1972,Kuenzle:1976,Ehlers:1981a,Ehlers:1981b}, \cite[chapter
4]{Malament:2012}, \cite{Hartong.EtAl:2023}.  A `teleparallellised'
version of Newton--Cartan gravity was recently constructed
\cite{Read.Teh:2018,Schwartz:2023}, which was shown to give an
analogous coordinate-free geometric description of the Newtonian limit
of \acronym{TEGR} \cite{Schwartz:2023} (as well as a description of
`null-reduced' \acronym{TEGR}, i.e.\ of the quotient of a solution of
\acronym{TEGR} by a null symmetry \cite{Read.Teh:2018}).  Very
recently, the Newtonian limit of \acronym{STEGR} was similarly given a
coordinate-free description based on Newton--Cartan gravity
\cite{Wolf.EtAl:2024}.

In the present paper, we give a full classification of affine
connections in Newton--Cartan geometry, i.e.\ on Galilei manifolds
(the `metric structure' underlying Newton--Cartan gravity), in terms
of freely specifiable tensor fields, analogous to the
pseudo-Riemannian case.  This generalises the well-known case of
(torsional) Galilei connections, i.e.\ connections compatible with the
metric structure, which is dealt with in, e.g., references
\cite{Bernal.Sanchez:2003,Bekaert.Morand:2016,Geracie.EtAl:2015,
  Bergshoeff.EtAl:2023}.  Note that special cases of this
classification have appeared and been used in reference
\cite{Wolf.EtAl:2024} already.  Here, however, we will give a full,
self-contained proof of the general result, and we are going to put
particular emphasis on which fields may be specified independently in
the definition of a connection.

Our result lays the foundation for studying the Newtonian limits of
general metric-affine theories of gravity in the language of
(modified) Newton--Cartan gravity, generalising and extending the
works mentioned above on the Newton--Cartan-like limits of
(\acronym{S})\acronym{TEGR} \cite{Read.Teh:2018,Schwartz:2023,
  Wolf.EtAl:2024}.  Only such a properly geometric, i.e.\
coordinate-free, formulation of the Newtonian limit can count as
providing a true understanding of a gravitational theory's Newtonian
behaviour: any formulation in concrete coordinates, as commonly
employed in the study of Newtonian limits and post-Newtonian
expansions, ignores the inherently geometric nature of gravity, which
is the main lesson of \acronym{GR}.  Building on such a geometric
understanding of the Newtonian limit, the results of the present paper
will enable the development of a geometric formulation of the
post-Newtonian expansion of general metric-affine theories, as has
been recently developed for standard \acronym{GR}
\cite{Dautcourt:1997,Tichy.Flanagan:2011,Van_den_Bleeken:2017,
  Hansen.EtAl:2019,Hansen.EtAl:2020,Hartong.EtAl:2023}.

The structure of this paper is as follows.  First, we introduce our
general notation and conventions in \cref{subsec:notation}.  In
\cref{sec:prelim}, we will recall the necessary background on Galilei
manifolds.  In \cref{sec:class}, we will formulate, discuss, and prove
our main result, namely the classification theorem for general affine
connections on Galilei manifolds.  Finally, in \cref{sec:bundle_persp}
we look at affine connections on Galilei manifolds from a different
angle, namely from the perspective of the principal bundle of Galilei
frames, before concluding in \cref{sec:conclusion}.

\subsection{Notation and conventions}
\label{subsec:notation}

All differentiable manifolds will be assumed smooth, i.e.\ $C^\infty$.
For the sake of simplicity, all objects on manifolds will also be
assumed smooth; note however that most of our statements will have
obvious generalisations to lower regularity.

We will denote the space of (smooth) sections of a fibre bundle $E \to
M$ by $\Gamma(E)$; the space of local sections on an open subset $U
\subset M$ is $\Gamma(U,E) := \Gamma(E|_U)$.  For the space of
differential $k$-forms on some manifold $M$, we will \emph{not} use
the notation $\Omega^k(M)$, but instead write it explicitly as
$\Gamma(\bigwedge^k T^*M)$.  The $k$-fold symmetric tensor power of a
vector space (or bundle) $V$ we will denote $\bigvee^k V$; so e.g.\
$\Gamma(\bigvee^2 T^*M)$ is the space of symmetric covariant
two-tensor fields on $M$.

By an \emph{affine connection} on a manifold $M$ we mean a covariant
derivative operator $\nabla$ on the tangent bundle $TM$, which we
extend to tensor bundles in the natural way, i.e.\ via demanding a
Leibniz rule.  For example, for vector fields $X, Y \in \Gamma(TM)$
and a one-form $\alpha \in \Gamma(T^*M)$, we have
\begin{equation}
  (\nabla_X \alpha)(Y) = X(\alpha(Y)) - \alpha(\nabla_X Y) \; .
\end{equation}
Connection coefficients in local coordinates are defined in the usual
way, i.e.\ by expanding the covariant derivatives of coordinate vector
fields according to
\begin{equation}
  \nabla_{\partial_\mu} \partial_\nu
  = \Gamma^\rho_{\mu\nu} \partial_\rho \; .
\end{equation}
Note that the \emph{first} lower index is the `form index' /
differentiation index, such that the coordinate component form of the
torsion of $\nabla$ is
\begin{equation}
  \tensor{T}{^\rho_{\mu\nu}} = 2 \Gamma^\rho_{[\mu\nu]} \; .
\end{equation}

In \cref{sec:bundle_persp}, we will deal with principal connections,
i.e.\ connections on principal bundles.  For those, we will use the
following convention: the (global) connection form living on the total
space of the bundle will be denoted by a boldface letter; local
connection forms on the base manifold, i.e.\ pullbacks along local
sections of the bundle, will be denoted by the corresponding
non-boldface letter, with the local section being understood from
context.  For example, given a connection form $\bomega \in
\Gamma(T^*P \otimes \mathfrak{g})$ on a principal $G$-bundle $P \to M$
(where $\mathfrak{g}$ is the Lie algebra of $G$), for a local section
$\sigma \in \Gamma(U, P)$ on an open subset $U \subset M$ the
corresponding local connection form is $\omega = \sigma^*\bomega \in
\Gamma(U, T^*M \otimes \mathfrak{g}) = \Gamma(T^*U \otimes
\mathfrak{g})$.

For tensorial forms on principal bundles, we will make an exception
from our rule of not notating spaces of differential forms with an
$\Omega$: given a representation $\rho \colon G \to \GL(V)$, the space
of $\rho$-tensorial $k$-forms on the total space $P$ of a principal
$G$-bundle will be denoted by $\Omega^k_\rho (P,V)$.

\section{Preliminaries}
\label{sec:prelim}

Here we recall the basic definitions regarding Galilei manifolds, as
given, e.g., by Künzle \cite{Kuenzle:1976}.  Note however that we use
a somewhat modernised notation, following, e.g., reference
\cite{Schwartz:2023}.

\begin{definition}
  A \emph{Galilei manifold} $(M,\tau,h)$ is a differentiable manifold
  $M$ of dimension $n+1$ (with $n \ge 1$) together with a
  nowhere-vanishing \emph{clock form} $\tau \in \Gamma(T^*M)$ and a
  \emph{space metric} $h \in \Gamma(\bigvee^2 TM)$ which is positive
  semidefinite of rank $n$, satisfying
  \begin{equation}
    \tau_\mu h^{\mu\nu} = 0 \; ,
  \end{equation}
  i.e.\ such that the degenerate direction of $h$ is spanned by
  $\tau$.

  Vectors in the kernel of $\tau$ are called \emph{spacelike}, other
  vectors are called \emph{timelike}.  A vector $v \in TM$ with
  $\tau(v) = 1$ is called \emph{unit timelike}.  We will mostly be
  concerned with (perhaps locally defined) unit timelike vector
  \emph{fields} $v \in \Gamma(U,TM), \tau(v) = 1$.

\end{definition}

With respect to a choice of a unit timelike vector field $v$, the
tangent and cotangent bundles of $M$ decompose as
\begin{subequations} \label{eq:decomp_time-space}
\begin{align}
  TM &= \mathrm{span}\{v\} \oplus \ker \tau \; , \\
  T^*M &= \mathrm{span}\{\tau\} \oplus \ker v \; .
\end{align}
\end{subequations}
Hence, $v$ induces a projector onto space and a kind of inverse to
$h$:
\begin{definition}
  Let $(M,\tau,h)$ be a Galilei manifold and $v$ a unit timelike
  vector field on it.
  \begin{enumerate}[label=(\roman*)]
  \item The \emph{spatial projector along $v$}, i.e.\ the unique
    projection operator $TM \to TM$ that projects onto $\ker\tau$ and
    whose kernel is spanned by $v$, will be denoted by $P$.
    Explicitly, it has components
    \begin{equation}
      P^\mu_\nu = \delta^\mu_\nu - v^\mu \tau_\nu \; .
    \end{equation}
    Note that we will not acknowledge the dependence of $P$ on $v$ in
    the notation.

  \item The \emph{covariant space metric with respect to $v$} is the
    unique symmetric covariant 2-tensor field $\hv = h_{\mu\nu} \, \D
    x^\mu \otimes \D x^\nu \in \Gamma(\bigvee^2 T^*M)$ defined by
    \begin{equation} \label{eq:def_hv}
      h_{\mu\nu} v^\nu = 0 \; , \quad
      h_{\mu\nu} h^{\nu\rho} = P^\rho_\mu \; .
    \end{equation}
    Note that in index notation, we leave out the subscript $v$ in
    order to avoid confusion with an index. An $h$ with indices
    `downstairs' will always mean the covariant space metric with
    respect to that unit timelike vector field $v$ which is clear from
    context.
  \end{enumerate}
\end{definition}

The decomposition of vectors and covectors according to
\eqref{eq:decomp_time-space} we call the decomposition into
\emph{timelike} and \emph{spacelike parts with respect to
  $v$}. Explicitly, for a vector $X$ and a covector $\alpha$, these
decompositions read
\begin{subequations}
\begin{align}
  X^\mu
  &= \delta^\mu_\nu X^\nu
    = v^\mu \tau_\nu X^\nu + P^\mu_\nu X^\nu , \\
  \alpha_\mu
  &= \delta^\nu_\mu \alpha_\nu
    =\tau_\mu v^\nu \alpha_\nu + P^\nu_\mu \alpha_\nu \; .
\end{align}
\end{subequations}

We will use the common convention of `raising indices' by contraction
with $h^{\mu\nu}$ and `lowering indices' by contraction with
$h_{\mu\nu}$ (if a unit timelike $v$ is clear from context).  Note
that due to the degeneracy of $h$ these operations are not inverses of
each other, but first raising and then lowering an index (or vice
versa) corresponds to contracting with $P^\mu_\nu$.

\section{The classification}
\label{sec:class}

In this section, we are going to state, discuss, and prove our main
result: the classification of affine connections on Galilei
manifolds in terms of free tensor fields.

First, we need to introduce a special tensor field associated with a
connection by choice of a unit timelike vector field.

\begin{definition}
  Let $(M,\tau,h)$ be a Galilei manifold, $v$ a unit timelike vector
  field on it, and $\nabla$ an affine connection on $M$.  The
  \emph{Newton--Coriolis form of $\nabla$ with respect to $v$} is the
  two-form $\Omega \in \Gamma(\bigwedge^2 T^*M)$ with components
  \begin{equation}
    \Omega_{\mu\nu} := 2 (\nabla_{[\mu} v^\rho) h_{\nu]\rho} \; .
  \end{equation}
\end{definition}

The name `Newton--Coriolis form', originally introduced in reference
\cite{Geracie.EtAl:2015}, comes from the case of standard
Newton--Cartan gravity.  There, the kinematics of a spacetime with
locally Newtonian physics is encoded into a torsion-free connection
$\nabla$ compatible with $\tau$ and $h$ (with $\D\tau = 0$), geodesics
of which are interpreted as the worldlines of freely falling
particles.  For a unit timelike vector field $v$, the spacelike fields
\begin{equation}
  \alpha^\mu := (\nabla_v v)^\mu \; , \quad
  \omega^{\mu\nu} := \nabla^{[\mu} v^{\nu]}
\end{equation}
can then be given the interpretation of the spacetime acceleration and
vorticity of $v$.  From the point of view of the reference system
defined by $v$, a freely falling particle with worldline $\gamma$ then
looks accelerated by a linear `gravitational' acceleration
$-\alpha^\mu$ and a Coriolis acceleration $2 \tensor{\omega}{^\mu_\nu}
\dot\gamma^\nu$.  The Newton--Coriolis form of $\nabla$ with respect
to $v$ can be decomposed as
\begin{equation} \label{eq:Omega_decomp}
  \Omega_{\mu\nu} = 2 \tau_{[\mu} \alpha_{\nu]} + 2 \omega_{\mu\nu} \; ,
\end{equation}
i.e.\ it combines the Newtonian gravity and Coriolis terms---hence the
name.

When leaving the realm of standard Newton--Cartan gravity and
considering general connections, the fields $\alpha$ and $\omega$
defined by the decomposition \eqref{eq:Omega_decomp} lose their
interpretation as acceleration and vorticity of $v$, and so the name
`Newton--Coriolis form' loses its justification.  Nevertheless, for
the sake of simplicity and continuity with the literature
\cite{Geracie.EtAl:2015,Schwartz:2023}, we are going to keep using
this name for $\Omega$ also in the cases of torsionful or `non-metric'
(i.e.\ incompatible with $\tau$ or $h$) connections.

Now we are in the position to formulate the general classification
result.
\begin{theorem} \label{thm:conn_class}
  Let $(M,\tau,h)$ be a Galilei manifold and $v$ a unit timelike
  vector field on it.

  \begin{enumerate}[label=(\roman*)]
  \item Let $\nabla$ be an affine connection on $M$.  Denote its
    torsion by $T$ and its Newton--Coriolis form with respect to $v$
    by $\Omega$, and define the `non-metricities' $\hat Q :=
    \nabla\tau \in \Gamma(T^*M \otimes T^*M)$ and $Q := \nabla h \in
    \Gamma(T^*M \otimes \bigvee^2 TM)$ (i.e.\ in components $\hat
    Q_{\mu\nu} := \nabla_\mu \tau_\nu$, $\tensor{Q}{_\rho^{\mu\nu}} :=
    \nabla_\rho h^{\mu\nu} = \tensor{Q}{_\rho^{\nu\mu}}$).  Then the
    identities
    \begin{equation} \label{eq:conn_class_ids}
      \tau_\mu \tensor{Q}{_\rho^{\mu\nu}}
      = -\hat Q_{\rho\mu} h^{\mu\nu} \; , \quad
      \tau_\rho \tensor{T}{^\rho_{\mu\nu}}
      = (\D\tau)_{\mu\nu} - 2 \hat Q_{[\mu\nu]}
    \end{equation}
    hold, and the connection coefficients of $\nabla$ take the form
    \begin{subequations} \label{eq:conn_class}
    \begin{align}
      \label{eq:conn_class_1}
      \Gamma^\rho_{\mu\nu}
      &= v^\rho \partial_\mu \tau_\nu
        + \frac{1}{2} h^{\rho\sigma} (\partial_\mu h_{\nu\sigma}
          + \partial_\nu h_{\mu\sigma} - \partial_\sigma h_{\mu\nu})
        + \frac{1}{2} P^\rho_\lambda \tensor{T}{^\lambda_{\mu\nu}}
        - \tensor{T}{_{(\mu\nu)}^\rho}
        + \tau_{(\mu} \tensor{\Omega}{_{\smash{\nu)}}^\rho}
        \nonumber\\
      &\quad- \frac{1}{2} \tensor{Q}{^\rho_{\mu\nu}}
        + P^\rho_\lambda \tensor{Q}{_{(\mu\nu)}^\lambda}
        - v^\rho \hat Q_{\mu\nu} \\
      \intertext{(where indices have been raised with $h$ and lowered
      with $\hv$).  Using the identities \eqref{eq:conn_class_ids},
      this may be rewritten as}
      \label{eq:conn_class_2}
      \Gamma^\rho_{\mu\nu}
      &= v^\rho \partial_{(\mu} \tau_{\nu)}
        + \frac{1}{2} h^{\rho\sigma} (\partial_\mu h_{\nu\sigma}
          + \partial_\nu h_{\mu\sigma} - \partial_\sigma h_{\mu\nu})
        + \frac{1}{2} \tensor{T}{^\rho_{\mu\nu}}
        - \tensor{T}{_{(\mu\nu)}^\rho}
        + \tau_{(\mu} \tensor{\Omega}{_{\smash{\nu)}}^\rho}
        \nonumber\\
      &\quad- \frac{1}{2} \tensor{Q}{^\rho_{\mu\nu}}
        + P^\rho_\lambda \tensor{Q}{_{(\mu\nu)}^\lambda}
        - v^\rho \hat Q_{(\mu\nu)} \\
      \label{eq:conn_class_3}
      &= v^\rho \partial_{(\mu} \tau_{\nu)}
        + \frac{1}{2} h^{\rho\sigma} (\partial_\mu h_{\nu\sigma}
          + \partial_\nu h_{\mu\sigma} - \partial_\sigma h_{\mu\nu})
        + \frac{1}{2} \tensor{T}{^\rho_{\mu\nu}}
        - \tensor{T}{_{(\mu\nu)}^\rho}
        + \tau_{(\mu} \tensor{\Omega}{_{\smash{\nu)}}^\rho}
        \nonumber\\
      &\quad- \frac{1}{2} \tensor{Q}{^\rho_{\mu\nu}}
        + \tensor{Q}{_{(\mu\nu)}^\rho}
        - v^\rho v^\sigma \tau_{(\mu} \hat Q_{\nu)\sigma} \; .
    \end{align}
    \end{subequations}

  \item \label{thm:conn_class_converse}
    Conversely, given arbitrary tensor fields $T \in \Gamma(TM
    \otimes \bigwedge^2 T^*M)$, $\hat Q \in \Gamma(T^*M \otimes T^*M)$
    and $Q \in \Gamma(T^*M \otimes \bigvee^2 TM)$ satisfying the
    identities \eqref{eq:conn_class_ids} and an arbitrary 2-form
    $\Omega \in \Gamma(\bigwedge^2 T^*M)$, the connection $\nabla$
    defined by \eqref{eq:conn_class} has torsion $T$ and satisfies
    $\nabla\tau = \hat Q$, $\nabla h = Q$, and has Newton--Coriolis
    form $\Omega$ with respect to $v$.
  \end{enumerate}
\end{theorem}

We wish to explicitly emphasize the `classification' point of view on
this result: \Cref{thm:conn_class} shows that (and how) a unit
timelike vector field $v$ induces a bijective correspondence between
affine connections on the one hand and tensor fields $T, \hat Q, Q,
\Omega$ satisfying \eqref{eq:conn_class_ids} (as well as their
respective symmetry properties) on the other hand.  We also note that
it generalises the well-known special cases of compatible torsion-free
connections as arising in standard Newton--Cartan gravity
\cite{Kuenzle:1972}, and of compatible torsionful connections
\cite{Bernal.Sanchez:2003,Bekaert.Morand:2016,Geracie.EtAl:2015,
  Bergshoeff.EtAl:2023}.

Before proving \cref{thm:conn_class}, we are going to give a few
remarks regarding the definition of a connection in terms of $T, \hat
Q, Q, \Omega$ according to part \ref{thm:conn_class_converse} (with
respect to a fixed $v$).
\begin{itemize}
\item Considering the torsion $T$, the identities
  \eqref{eq:conn_class_ids} restrict only its timelike part on the
  first index $\tau_\rho \tensor{T}{^\rho_{\mu\nu}}$, leaving the
  spacelike part $P^\rho_\lambda \tensor{T}{^\lambda_{\mu\nu}}$
  (equivalently $T_{\lambda\mu\nu}$) unconstrained, i.e.\ freely
  specifiable.

\item Similarly, \eqref{eq:conn_class_ids} restrict of the
  $h$-non-metricity $Q = \nabla h$ only the timelike part on the rear
  indices $\tau_\mu \tensor{Q}{_\rho^{\mu\nu}} = \tau_\mu
  \tensor{Q}{_\rho^{\nu\mu}}$, leaving the spacelike part
  $\tensor{Q}{_\rho^{\kappa\lambda}} P^\mu_\kappa P^\nu_\lambda$
  (equivalently $Q_{\rho\kappa\lambda}$) freely specifiable.

\item The restricted parts $\tau_\rho \tensor{T}{^\rho_{\mu\nu}}$ of
  the torsion and $\tau_\mu \tensor{Q}{_\rho^{\mu\nu}}$ of the
  $h$-non-metricity $Q = \nabla h$ are determined by
  \eqref{eq:conn_class_ids} in terms of the $\tau$-non-metricity $\hat
  Q = \nabla\tau$, which may be specified arbitrarily.  From this
  point of view, a choice of unit timelike vector field $v$ therefore
  induces an affine isomorphism between the affine space of affine
  connections on $M$ and the vector space
  \begin{subequations} \label{eq:conn_free_fields}
  \begin{equation} \label{eq:conn_free_fields_space} \textstyle
    \Gamma(\ker\tau \otimes \bigwedge^2 T^*M) \oplus
    \Gamma(T^*M \otimes T^*M) \oplus
    \Gamma(T^*M \otimes \bigvee^2 \ker\tau) \oplus
    \Gamma(\bigwedge^2 T^*M) \; ,
  \end{equation}
  mapping a connection $\nabla$ to the field with components
  \begin{equation} \label{eq:conn_free_fields_comp}
    (P^\rho_\lambda \tensor{T}{^\lambda_{\mu\nu}} \, ,
    \hat Q_{\mu\nu} \, ,
    \tensor{Q}{_\rho^{\kappa\lambda}} P^\mu_\kappa P^\nu_\lambda \, ,
    \Omega_{\mu\nu}) \; .
  \end{equation}
  \end{subequations}
  The first form \eqref{eq:conn_class_1} of the connection
  coefficients from the theorem expresses the connection explicitly in
  terms of only these freely specifiable fields, i.e.\ it provides
  the inverse of this isomorphism explicitly.

  This is the direct generalisation of theorem~5.27 of reference
  \cite{Bernal.Sanchez:2003} and proposition~4.3 / theorem~4.5 of
  reference \cite{Bekaert.Morand:2016} to the non-metric case.

\item Using \eqref{eq:conn_free_fields}, we can count the number of
  pointwise algebraically independent components of a general
  connection arising from each of the individual `pieces' in the
  classification, i.e.\ the dimensions of the fibres of the respective
  vector bundles in \eqref{eq:conn_free_fields_space}: writing $\dim M
  = n + 1$, at each point the Newton--Coriolis form has
  \begin{subequations}
  \begin{align}
    \textstyle \dim(\bigwedge^2 T_p^*M)
    &= \frac{n(n+1)}{2} \\
    \intertext{independent components, the spacelike part of the
    torsion has}
    \textstyle \dim(\ker\tau|_p \otimes \bigwedge^2 T_p^*M)
    &= n \cdot \frac{n(n+1)}{2} = \frac{n^2(n+1)}{2} \; ,\\
    \intertext{and the non-metricities have}
    \textstyle \dim\big((T_p^*M \otimes T_p^*M)
      \oplus (T_p^*M \otimes \bigvee^2 \ker\tau|_p)\big)
    &= (n+1)^2 + (n+1) \cdot \frac{n(n+1)}{2} \nonumber\\
    &= \frac{(n+2)(n+1)^2}{2} \; .
  \end{align}
  \end{subequations}
  In particular, the affine bundle of \emph{compatible} connections
  (allowing for torsion) has rank $\frac{n(n+1)}{2} +
  \frac{n^2(n+1)}{2} = \frac{(n+n^2)(n+1)}{2} = \frac{n(n+1)^2}{2}$.
  Adding to this the pointwise dimension of the space of possible
  non-metricities, we obtain $\frac{n(n+1)^2}{2} +
  \frac{(n+2)(n+1)^2}{2} = \frac{(2n+2)(n+1)^2}{2} = (n+1)^3$---as we
  have to, since this is the number of pointwise independent
  components of a general affine connection on a manifold of dimension
  $n+1$.

\item Consider the connection corresponding to the zero field in
  \eqref{eq:conn_free_fields_space} under the isomorphism induced by
  $v$, i.e.\ the connection determined by having (i) vanishing
  `spatial torsion' with respect to $v$, (ii) vanishing
  $\tau$-non-metricity $\hat Q$, (iii) vanishing `spatial
  $h$-non-metricity' with respect to $v$, and (iv) vanishing
  Newton--Coriolis form with respect to $v$.  By
  \eqref{eq:conn_class_ids}, we then have $Q = 0$ as well, i.e.\ the
  connection is a Galilei connection (compatible with $\tau$ and $h$),
  and its torsion is $\tensor{T}{^\rho_{\mu\nu}} = v^\rho
  (\D\tau)_{\mu\nu}$.  This is what is commonly called the `special
  Galilei connection' with respect to $v$.

\item Let us stress explicitly that the unit timelike vector field $v$
  is \emph{not} an independent field in defining an affine connection.
  On the contrary, a choice of $v$ only determines an affine
  isomorphism between the affine space of affine connections and the
  vector space \eqref{eq:conn_free_fields_space}; the freedom in
  specifying the connection itself resides (from this perspective)
  fully in this vector space.

  Put differently: once a fixed connection $\nabla$ is being
  considered, it may be represented with respect to \emph{any} choice
  of unit timelike vector field $v$; under a change of $v$ (called a
  `Milne boost' or `local Galilei boost' in the literature), the field
  \eqref{eq:conn_free_fields_comp} representing $\nabla$ with respect
  to $v$ will change accordingly.  This means that affine connections
  on $M$ correspond to Milne boost equivalence classes $[v,
  (P^\rho_\lambda \tensor{T}{^\lambda_{\mu\nu}} \, , \hat Q_{\mu\nu}
  \, , \tensor{Q}{_\rho^{\kappa\lambda}} P^\mu_\kappa P^\nu_\lambda \,
  , \Omega_{\mu\nu})]$ of pairs of unit timelike vector fields $v$ and
  fields in \eqref{eq:conn_free_fields_space}.

\item Note that torsion and non-metricity not being independent is in
  distinction to the pseudo-Riemannian case.  Hence, differently to
  the pseudo-Riemannian case, we cannot meaningfully separate the
  difference of a general affine connection to a metric-compatible
  torsion-free one into a torsion-dependent `contortion' and a
  non-metricity-dependent `distortion' part: even when (a) assuming
  that a metric-compatible torsion-free connection exists (necessarily
  implying `absolute time', i.e.\ $\D\tau = 0$), and (b) ignoring the
  additional freedom in $\Omega$, we can always trade some torsion for
  non-metricity and vice versa by virtue of the second identity of
  \eqref{eq:conn_class_ids}.

\item In applications of (Lorentzian) metric-affine geometry in the
  construction of gravitational theories, a major role is played by
  the connection's curvature.  Also in standard Newton--Cartan
  gravity, the gravitational field equation (which `geometrises' the
  Poisson equation of standard Newtonian gravity) is formulated in
  terms of the Ricci tensor of the connection (which is compatible
  with $\tau$ and $h$).  Hence it is natural to expect that the
  curvature will also enter in an integral way the formulation of
  metric-affine variants of Newton--Cartan gravity, which provide the
  Newtonian limit of metric-affine theories of gravity.

  The expression \eqref{eq:conn_class} for a general connection
  $\nabla$ on a Galilei manifold in terms of the independent fields in
  \eqref{eq:conn_free_fields_space} enables a direct computation of
  the curvature tensor of $\nabla$ in terms of these fields, which
  will probably be of importance in the investigation of gravitational
  theories.  Furthermore, the curvature will of course satisfy the
  Bianchi identities, which in the general case (i.e.\ with torsion)
  take the component form
  \begin{equation}
    \tensor{R}{^\mu_{[\nu\rho\sigma]}}
    = \nabla^{\vphantom{\mu}}_{[\rho} \tensor{T}{^\mu_{\sigma\nu]}}
    + \tensor{T}{^\mu_{\kappa[\nu}}
    \tensor{T}{^{\kappa\vphantom{\mu}}_{\rho\sigma]}} \; , \quad
    \nabla^{\vphantom{\mu}}_{[\kappa}
    \tensor{R}{^\mu_{|\nu|\rho\sigma]}}
    + \tensor{R}{^\mu_{\nu\lambda[\kappa}}
    \tensor{T}{^{\lambda\vphantom{\mu}}_{\rho\sigma]}}
    = 0 \; .
  \end{equation}
  However, spelling out the explicit form of both the curvature tensor
  and the Bianchi identities in terms of the independent fields
  \eqref{eq:conn_free_fields_comp} defining the connection we leave to
  future work.
\end{itemize}

Now we are going to prove the classification theorem.  Note that many
of the steps of the proof are slight generalisations of the
corresponding arguments in the well-known metric-compatible case.  We
decided, however, not to take the latter for granted, but present here
a fully detailed, self-contained proof, in order to (a) show how the
full classification formula emerges naturally, and (b) not have the
proof scattered across the literature.

\begin{proof}[Proof of \cref{thm:conn_class}]
  \begin{enumerate}[label=(\roman*)]
  \item The first identity of \eqref{eq:conn_class_ids} follows
    directly by taking the covariant derivative of the equation
    $\tau_\mu h^{\mu\nu} = 0$.  The second we state as
    \begin{lemma} \label{lem:conn_temporal_torsion}
      Let $(M,\tau,h)$ be a Galilei manifold and $\nabla$ an affine
      connection on $M$.  Then the torsion of $\nabla$ satisfies
      \begin{subequations}
      \begin{equation}
        \tau(T(X,Y))
        = \D\tau(X,Y) - (\nabla_X\tau)(Y) + (\nabla_Y\tau)(X)
      \end{equation}
      for any two vector fields $X,Y$, i.e.\
      \begin{equation}
        \tau_\rho \tensor{T}{^\rho_{\mu\nu}}
        = (\D\tau)_{\mu\nu} - 2 \nabla_{[\mu} \tau_{\nu]}
      \end{equation}
      \end{subequations}
      in index notation.

      \begin{proof}
        Direct computation using the invariant formula for the
        exterior derivative yields
        \begin{align}
          \tau(T(X,Y)) 
          &= \tau(\nabla_X Y) - \tau(\nabla_Y X) - \tau([X,Y])
            \nonumber\\
          &= X(\tau(Y)) - (\nabla_X\tau)(Y)
            - Y(\tau(X)) + (\nabla_Y\tau)(X)
            - \tau([X,Y]) \nonumber\\
          &= \D\tau(X,Y) - (\nabla_X\tau)(Y) + (\nabla_Y\tau)(X).
        \end{align}
      \end{proof}
    \end{lemma}

    Now we turn to the derivation of the classification formula
    \eqref{eq:conn_class}.  We decompose the connection coefficients
    on their upper index into their timelike and spacelike parts with
    respect to $v$, according to
    \begin{equation}
      \Gamma^\rho_{\mu\nu}
      = \delta^\rho_\lambda \Gamma^\lambda_{\mu\nu}
      = (v^\rho \tau_\lambda + P^\rho_\lambda) \Gamma^\lambda_{\mu\nu}
      = (v^\rho \tau_\lambda + h^{\rho\sigma} h_{\sigma\lambda})
        \Gamma^\lambda_{\mu\nu} \; .
    \end{equation}
    The timelike part we obtain directly from $\hat Q_{\mu\nu} =
    \nabla_\mu \tau_\nu = \partial_\mu \tau_\nu -
    \Gamma^\lambda_{\mu\nu} \tau_\lambda$, giving
    \begin{equation} \label{eq:class_conn_pf_coeff_decomp}
      \Gamma^\rho_{\mu\nu}
      = v^\rho \partial_\mu \tau_\nu - v^\rho \hat Q_{\mu\nu}
        + h^{\rho\sigma} h_{\sigma\lambda} \Gamma^\lambda_{\mu\nu}
      \; .
    \end{equation}
    To compute the spacelike part, we will proceed in quite some
    parallel to the standard way of classifying affine connections on
    pseudo-Riemannian manifolds.

    For any three vector fields $X,Y,Z$, by applying the Leibniz rule
    and the definition of the torsion we have
    \begin{subequations}
    \begin{align}
      \label{eq:class_conn_pf_spacel_1}
      \hv(\nabla_X Y, Z)
      &= X \big(\hv(Y,Z)\big)
        - \big(\nabla_X \hv\big)(Y,Z)
        - \hv(Y, \underbrace{\nabla_X Z}_{\mathrlap{= \nabla_Z X
          - [Z,X] - T(Z,X)}}) \nonumber\\
      &= X \big(\hv(Y,Z)\big) - \big(\nabla_X \hv\big)(Y,Z)
        - \hv(Y, \nabla_Z X) + \hv(Y, T(Z,X) + [Z,X]) \; . \\
      \intertext{Cyclicly permuting the vector fields, from this
      we obtain}
      \label{eq:class_conn_pf_spacel_2}
      \hv(\nabla_Y Z, X)
      &= Y \big(\hv(Z,X)\big) - \big(\nabla_Y \hv\big)(Z,X)
        - \hv(Z, \nabla_X Y) + \hv(Z, T(X,Y) + [X,Y]) \; , \\
      \label{eq:class_conn_pf_spacel_3}
      \hv(\nabla_Z X, Y)
      &= Z \big(\hv(X,Y)\big) - \big(\nabla_Z \hv\big)(X,Y)
        - \hv(X, \nabla_Y Z) + \hv(X, T(Y,Z) + [Y,Z]) \; .
    \end{align}
    \end{subequations}
    Considering $\eqref{eq:class_conn_pf_spacel_1} +
    \eqref{eq:class_conn_pf_spacel_2} -
    \eqref{eq:class_conn_pf_spacel_3}$ now yields
    \begin{align} \label{eq:class_conn_pf_spacel_sum}
      2 \hv(\nabla_X Y, Z)
      &= X \big(\hv(Y,Z)\big) + Y \big(\hv(Z,X)\big)
        - Z \big(\hv(X, Y)\big) \nonumber\\
      &\quad- \big(\nabla_X \hv\big)(Y,Z) - \big(\nabla_Y\hv\big)(Z,X)
        + \big(\nabla_Z \hv\big)(X,Y) \nonumber\\
      &\quad+ \hv(Y, T(Z,X) + [Z,X]) + \hv(Z, T(X,Y) + [X,Y])
        \nonumber\\
      &\quad- \hv(X,T(Y,Z) + [Y,Z]) \; .
    \end{align}
    To continue, we compute how to to express the covariant derivative
    $\nabla\hv$ of the covariant space metric in terms of $\nabla h$
    and $\nabla v$, for which we introduce the shorthand notation
    $\tensor{\Lambda}{_\mu^\nu} := \nabla_\mu v^\nu$:
    \begin{lemma} \label{lem:cov_der_hv}
      Let $(M,\tau,h)$ be a Galilei manifold, and consider a unit
      timelike vector field $v$ and an affine connection $\nabla$ on
      it.  Writing $\tensor{\Lambda}{_\mu^\nu} = \nabla_\mu v^\nu$ and
      $\tensor{Q}{_\rho^{\mu\nu}} = \nabla_\rho h^{\mu\nu}$, the
      covariant derivative of the covariant space metric $\hv$ is
      \begin{equation} \label{eq:cov_der_hv}
        \nabla_\rho h_{\mu\nu}
        = - 2 \Lambda_{\rho(\mu} \tau_{\nu)} - Q_{\rho\mu\nu}
      \end{equation}
      (where indices have been lowered with $\hv$).

      \begin{proof}
        We decompose $\nabla\hv$ on its second index into its timelike
        and its spacelike part with respect to $v$, namely
        \begin{equation} \label{eq:cov_der_hv_decomp}
          \nabla_\rho h_{\mu\nu}
          = \tau_\mu v^\lambda \nabla_\rho h_{\lambda\nu}
            + h_{\mu\kappa} h^{\kappa\lambda}
              \nabla_\rho h_{\lambda\nu} \; .
        \end{equation}
        The timelike part we obtain from
        \begin{subequations}
        \begin{equation}
          0 = \nabla_\rho (v^\lambda h_{\lambda\nu})
          = \tensor{\Lambda}{_\rho^\lambda} h_{\lambda\nu}
            + v^\lambda \nabla_\rho h_{\lambda\nu}
          = \Lambda_{\rho\nu} + v^\lambda \nabla_\rho h_{\lambda\nu}
          \; ,
        \end{equation}
        and the spacelike one from
        \begin{align}
          0 &= \nabla_\rho \delta^\kappa_\nu
              = \nabla_\rho (P^\kappa_\nu + v^\kappa \tau_\nu)
              \nonumber\\
            &= \nabla_\rho (h^{\kappa\lambda} h_{\lambda\nu})
              + \tensor{\Lambda}{_\rho^\kappa} \tau_\nu
              + v^\kappa \nabla_\rho \tau_\nu \nonumber\\
            &= h^{\kappa\lambda} \nabla_\rho h_{\lambda\nu}
              + \tensor{Q}{_\rho^{\kappa\lambda}} h_{\lambda\nu}
              + \tensor{\Lambda}{_\rho^\kappa} \tau_\nu
              + v^\kappa \nabla_\rho \tau_\nu  \nonumber\\
            &= h^{\kappa\lambda} \nabla_\rho h_{\lambda\nu}
              + \tensor{Q}{_\rho^\kappa_\nu}
              + \tensor{\Lambda}{_\rho^\kappa} \tau_\nu
              + v^\kappa \nabla_\rho \tau_\nu \; .
        \end{align}
        \end{subequations}
        Inserting these into \eqref{eq:cov_der_hv_decomp}, we obtain
        \eqref{eq:cov_der_hv}.
      \end{proof}
    \end{lemma}

    Applying \eqref{eq:class_conn_pf_spacel_sum} to the coordinate
    vector fields $X = \partial_\mu$, $Y = \partial_\nu$, $Z =
    \partial_\sigma$ and using \cref{lem:cov_der_hv} to express
    $\nabla\hv$ in terms of $\Lambda = \nabla v$ and $Q = \nabla h$,
    we obtain
    \begin{align} \label{eq:class_Galilei_conn_pf_coeff_spacel}
      h_{\sigma\lambda} \Gamma^\lambda_{\mu\nu}
      &= \frac{1}{2} (\partial_\mu h_{\nu\sigma}
          + \partial_\nu h_{\sigma\mu} - \partial_\sigma h_{\mu\nu})
        + \frac{1}{2} (-\nabla_\mu h_{\nu\sigma}
          - \nabla_\nu h_{\sigma\mu} + \nabla_\sigma h_{\mu\nu})
        \nonumber\\
      &\quad+ \frac{1}{2} (T_{\nu\sigma\mu} + T_{\sigma\mu\nu}
          - T_{\mu\nu\sigma}) \nonumber\\
      &= \frac{1}{2} (\partial_\mu h_{\nu\sigma}
          + \partial_\nu h_{\sigma\mu} - \partial_\sigma h_{\mu\nu})
        + \Lambda_{\mu(\nu} \tau_{\sigma)}
        + \Lambda_{\nu(\sigma} \tau_{\mu)}
        - \Lambda_{\sigma(\mu} \tau_{\nu)} \nonumber\\
      &\quad+ \frac{1}{2} (Q_{\mu\nu\sigma} + Q_{\nu\sigma\mu}
          - Q_{\sigma\mu\nu})
        + \frac{1}{2} (T_{\nu\sigma\mu} + T_{\sigma\mu\nu}
          - T_{\mu\nu\sigma}) \nonumber\displaybreak[0]\\
      &= \frac{1}{2} (\partial_\mu h_{\nu\sigma}
          + \partial_\nu h_{\sigma\mu} - \partial_\sigma h_{\mu\nu})
        + \Lambda_{\mu(\nu} \tau_{\sigma)}
        + \Lambda_{\nu(\sigma} \tau_{\mu)}
        - \Lambda_{\sigma(\mu} \tau_{\nu)} \nonumber\\
      &\quad- \frac{1}{2} Q_{\sigma\mu\nu} + Q_{(\mu\nu)\sigma}
        + \frac{1}{2} T_{\sigma\mu\nu} - T_{(\mu\nu)\sigma} \; .
    \end{align}
    In the last step, we have used the symmetry properties of $T$ and
    $Q$ to combine some terms.

    Equation \eqref{eq:class_Galilei_conn_pf_coeff_spacel} gives us
    the desired expression for the spacelike part of the connection
    coefficients.  Inserting it into
    \eqref{eq:class_conn_pf_coeff_decomp}, we obtain
    \begin{align} \label{eq:class_conn_pf_coeff}
      \Gamma^\rho_{\mu\nu}
      &= v^\rho \partial_\mu \tau_\nu
        + \frac{1}{2} h^{\rho\sigma}
          (\partial_\mu h_{\nu\sigma} + \partial_\nu h_{\mu\sigma}
          - \partial_\sigma h_{\mu\nu})
        + \frac{1}{2} h^{\rho\sigma} T_{\sigma\mu\nu}
        - h^{\rho\sigma} T_{(\mu\nu)\sigma} \nonumber\\
      &\quad+ h^{\rho\sigma}
          (\Lambda_{\mu(\nu} \tau_{\sigma)}
          + \Lambda_{\nu(\sigma} \tau_{\mu)}
          - \Lambda_{\sigma(\mu} \tau_{\nu)})
        - \frac{1}{2} h^{\rho\sigma} Q_{\sigma\mu\nu}
        + h^{\rho\sigma} Q_{(\mu\nu)\sigma}
        - v^\rho \hat Q_{\mu\nu} \nonumber\\
      &= v^\rho \partial_\mu \tau_\nu
        + \frac{1}{2} h^{\rho\sigma}
          (\partial_\mu h_{\nu\sigma} + \partial_\nu h_{\mu\sigma}
          - \partial_\sigma h_{\mu\nu})
        + \frac{1}{2} P^\rho_\lambda \tensor{T}{^\lambda_{\mu\nu}}
        - \tensor{T}{_{(\mu\nu)}^\rho} \nonumber\\
      &\quad+ h^{\rho\sigma} (\Lambda_{(\mu|\sigma|}
          - \Lambda_{\sigma(\mu}) \tau_{\nu)}
        - \frac{1}{2} \tensor{Q}{^\rho_{\mu\nu}}
        + P^\rho_\lambda \tensor{Q}{_{(\mu\nu)}^\lambda}
        - v^\rho \hat Q_{\mu\nu} \; ,
    \end{align}
    where we used that first lowering an index with $\hv$ and then
    raising it again with $h$ amounts to spatially projecting it with
    $P$.  Noting that the definition of the Newton--Coriolis form is
    $\Omega_{\mu\rho} = 2 \Lambda_{[\mu\rho]} = \Lambda_{\mu\rho} -
    \Lambda_{\rho\mu}$, we may rewrite
    \begin{equation}
      (\Lambda_{(\mu|\sigma|} - \Lambda_{\sigma(\mu}) \tau_{\nu)}
      = \Omega_{(\mu|\sigma|} \tau_{\nu)}
      = \tau_{(\mu} \Omega_{\nu)\sigma} \; .
    \end{equation}
    Thus, \eqref{eq:class_conn_pf_coeff} yields the connection
    coefficients as stated in the first version of the classification
    formula \eqref{eq:conn_class_1}.

    Finally, we are going to show that due to the identities
    \eqref{eq:conn_class_ids}, the three forms
    \eqref{eq:conn_class_1}, \eqref{eq:conn_class_2}, and
    \eqref{eq:conn_class_3} of the connection coefficients are equal.
    First considering the spatially projected torsion term in
    \eqref{eq:conn_class_1}, using the second identity of
    \eqref{eq:conn_class_ids} it can be rewritten as
    \begin{align} \label{eq:class_conn_pf_torsion}
      \frac{1}{2} P^\rho_\lambda \tensor{T}{^\lambda_{\mu\nu}}
      &= \frac{1}{2} \tensor{T}{^\rho_{\mu\nu}}
        - \frac{1}{2} v^\rho \tau_\lambda
          \tensor{T}{^\lambda_{\mu\nu}} \nonumber\\
      &= \frac{1}{2} \tensor{T}{^\rho_{\mu\nu}}
        - \frac{1}{2} v^\rho (\D\tau)_{\mu\nu}
        + v^\rho \nabla_{[\mu} \tau_{\nu]} \nonumber\\
      &= \frac{1}{2} \tensor{T}{^\rho_{\mu\nu}}
        - v^\rho \partial_{[\mu} \tau_{\nu]}
        + v^\rho \hat Q_{[\mu\nu]} \; .
    \end{align}
    This shows that \eqref{eq:conn_class_1} and
    \eqref{eq:conn_class_2} are indeed equal.
    
    Secondly, considering the spatially projected non-metricity term
    in \eqref{eq:conn_class_1}/\eqref{eq:conn_class_2}, writing out
    the spatial projector we have $P^\rho_\lambda
    \tensor{Q}{_{(\mu\nu)}^\lambda} = \tensor{Q}{_{(\mu\nu)}^\rho} -
    v^\rho \tau_\lambda \tensor{Q}{_{(\mu\nu)}^\lambda}$.  For the
    second term in this expression, the first identity of
    \eqref{eq:conn_class_ids} yields
    \begin{equation} \label{eq:conn_class_id_lower}
      \tau_\lambda \tensor{Q}{_{\mu\nu}^\lambda}
      = \tau_\lambda \tensor{Q}{_\mu^{\sigma\lambda}} h_{\nu\sigma}
      = - \hat Q_{\mu\lambda} h^{\sigma\lambda} h_{\nu\sigma}
      = - \hat Q_{\mu\lambda} P^\lambda_\nu
      = - \hat Q_{\mu\nu} + v^\lambda \tau_\nu \hat Q_{\mu\lambda}
      \; .
    \end{equation}
    Inserting this, we have
    \begin{equation} \label{eq:class_conn_pf_non-metr}
      P^\rho_\lambda \tensor{Q}{_{(\mu\nu)}^\lambda}
      = \tensor{Q}{_{(\mu\nu)}^\rho} + v^\rho \hat Q_{(\mu\nu)}
      - v^\rho v^\lambda \tau_{(\mu} \hat Q_{\nu)\lambda} \; ,
    \end{equation}
    which shows equality of \eqref{eq:conn_class_2} and
    \eqref{eq:conn_class_3}.

  \item We now turn to the proof of the converse, i.e.\ that the
    connection $\nabla$ defined by \eqref{eq:conn_class} for given $T,
    Q, \hat Q, \Omega$ has torsion $T$, satisfies $\nabla\tau = \hat
    Q$, $\nabla h = Q$, and has Newton--Coriolis form $\Omega$ with
    respect to $v$.

    First, by antisymmetrising the second form \eqref{eq:conn_class_2}
    of the connection coefficients in its lower indices, we directly
    obtain $2\Gamma^\rho_{[\mu\nu]} = \tensor{T}{^\rho_{\mu\nu}}$,
    showing that the torsion of $\nabla$ is indeed $T$.

    Next, contracting the first form \eqref{eq:conn_class_1} of the
    connection coefficients with $\tau_\rho$, most terms vanish since
    $\rho$ was raised in them with $h$, leaving
    \begin{equation}
      \tau_\rho \Gamma^\rho_{\mu\nu}
      = \partial_\mu \tau_\nu - \hat Q_{\mu\nu} \; .
    \end{equation}
    This shows $\nabla_\mu \tau_\nu = \partial_\mu \tau_\nu -
    \Gamma^\rho_{\mu\nu} \tau_\rho = \hat Q_{\mu\nu}$.

    In order to compute $\nabla h$ and the Newton--Coriolis form of
    $\nabla$ with respect to $v$, we use the following standard
    result, a proof of which may be found in \cref{sec:app_spec_conn}:
    \begin{lemma} \label{lem:nablav}
      Let $(M,\tau,h)$ be a Galilei manifold and $v$ a unit timelike
      vector field on it.  The connection $\nablav$ defined by having
      coefficients
      \begin{equation} \label{eq:Gammav}
        \Gammav^\rho_{\mu\nu}
        = v^\rho \partial_\mu \tau_\nu
          + \frac{1}{2} h^{\rho\sigma} (\partial_\mu h_{\nu\sigma}
            + \partial_\nu h_{\mu\sigma} - \partial_\sigma h_{\mu\nu})
      \end{equation}
      is a Galilei connection, i.e.\ satisfies $\nablav\tau = 0$,
      $\nablav h = 0$, and its Newton--Coriolis form with respect to
      $v$ vanishes.
    \end{lemma}

    Denoting the difference tensor between $\nabla$ and $\nablav$ by
    $\tensor{S}{^\rho_{\mu\nu}} := \Gamma^\rho_{\mu\nu} -
    \Gammav^\rho_{\mu\nu}$, by \cref{lem:nablav} for $\nabla h$ we
    have
    \begin{equation} \label{eq:non-metr_via_diff}
      \nabla_\mu h^{\rho\sigma}
      = \nablav_\mu h^{\rho\sigma}
        + \tensor{S}{^\rho_{\mu\lambda}} h^{\lambda\sigma}
        + \tensor{S}{^\sigma_{\mu\lambda}} h^{\rho\lambda}
      = 2 \tensor{S}{^{(\rho}_\mu^{\sigma)}} \; ,
    \end{equation}
    and the Newton--Coriolis form $\tilde{\Omega}$ of $\nabla$ with
    respect to $v$ has components
    \begin{equation} \label{eq:NC_form_via_diff}
      \tilde{\Omega}_{\mu\rho}
      = 2 (\nabla_{[\mu} v^\kappa) h_{\rho]\kappa}
      = 2 (\nablav_{[\mu} v^\kappa
        + \tensor{S}{^\kappa_{[\mu|\nu|}} v^\nu) h_{\rho]\kappa}
      = 2 S_{[\rho\mu]\nu} v^\nu \; .
    \end{equation}
    Note that we will denote the Newton--Coriolis form by
    $\tilde{\Omega}$, since $\Omega$ refers to the given two-form that
    went into the definition of $\nabla$ (and of which we are to show
    that it equals $\tilde{\Omega}$).

    From \eqref{eq:conn_class_1} and \eqref{eq:Gammav}, the explicit
    form of the difference tensor is
    \begin{equation} \label{eq:nabla-nablav}
      \tensor{S}{^\rho_{\mu\nu}}
      = \frac{1}{2} P^\rho_\lambda \tensor{T}{^\lambda_{\mu\nu}}
        - \tensor{T}{_{(\mu\nu)}^\rho}
        + \tau_{(\mu} \tensor{\Omega}{_{\smash{\nu)}}^\rho}
        - \frac{1}{2} \tensor{Q}{^\rho_{\mu\nu}}
        + P^\rho_\lambda \tensor{Q}{_{(\mu\nu)}^\lambda}
        - v^\rho \hat Q_{\mu\nu} \; .
    \end{equation}

    Raising the third index of \eqref{eq:nabla-nablav}, we obtain
    \begin{align}
      2 \tensor{S}{^\rho_\mu^\sigma}
      &= P^\rho_\lambda \tensor{T}{^\lambda_\mu^\sigma}
        - \tensor{T}{_\mu^{\sigma\rho}}
        - P^\sigma_\lambda \tensor{T}{^\lambda_\mu^\rho}
        + \tau_\mu \Omega^{\sigma\rho} \nonumber\\
      &\quad- P^\sigma_\lambda \tensor{Q}{^\rho_\mu^\lambda}
        + P^\rho_\lambda P^\sigma_\kappa
          \tensor{Q}{_\mu^{\kappa\lambda}}
        + P^\rho_\lambda \tensor{Q}{^\sigma_\mu^\lambda}
        - 2 v^\rho \tensor{\hat Q}{_\mu^\sigma} \; .
    \end{align}
    Most of the terms in this expression are antisymmetric under
    exchange of $\rho$ and $\sigma$, so symmetrising yields
    \begin{equation}
      2 \tensor{S}{^{(\rho}_\mu^{\sigma)}}
      = P^\rho_\kappa P^\sigma_\lambda
          \tensor{Q}{_\mu^{\kappa\lambda}}
        - 2 v^{(\rho} \tensor{\hat Q}{_\mu^{\smash{\sigma)}}} \; .
    \end{equation}
    Writing out the spatial projectors and applying the first identity
    of \eqref{eq:conn_class_ids}, we have
    \begin{align}
      2 \tensor{S}{^{(\rho}_\mu^{\sigma)}}
      &= P^\rho_\kappa \tensor{Q}{_\mu^{\kappa\sigma}}
        - P^\rho_\kappa v^\sigma \underbrace{\tau_\lambda
            \tensor{Q}{_\mu^{\kappa\lambda}}}_{=
            -\tensor{\hat Q}{_\mu^\kappa}}{}
        - 2 v^{(\rho} \tensor{\hat Q}{_\mu^{\smash{\sigma)}}}
        \nonumber\\
      &= \tensor{Q}{_\mu^{\rho\sigma}}
        - v^\rho \underbrace{\tau_\kappa
            \tensor{Q}{_\mu^{\kappa\sigma}}}_{=
            -\tensor{\hat Q}{_\mu^\sigma}}{}
        + v^\sigma \tensor{\hat Q}{_\mu^\rho}
        - 2 v^{(\rho} \tensor{\hat Q}{_\mu^{\smash{\sigma)}}}
        \nonumber\\
      &= \tensor{Q}{_\mu^{\rho\sigma}} \; ,
    \end{align}
    which together with \eqref{eq:non-metr_via_diff} shows $\nabla h =
    Q$.

    Lowering the first index of the difference tensor
    \eqref{eq:nabla-nablav}, we obtain
    \begin{equation}
      S_{\rho\mu\nu}
      = \frac{1}{2} T_{\rho\mu\nu}
        - P^\lambda_\rho T_{(\mu\nu)\lambda}
        + P^\lambda_\rho \tau_{(\mu} \Omega_{\nu)\lambda}
        - \frac{1}{2} P^\lambda_\rho Q_{\lambda\mu\nu}
        + Q_{(\mu\nu)\rho} \; .
    \end{equation}
    Contracting this with $2 v^\nu$ and writing out the spatial
    projectors yields
    \begin{align}
      2 S_{\rho\mu\nu} v^\nu
      &= T_{\rho\mu\nu} v^\nu
        - P^\lambda_\rho T_{\mu\nu\lambda} v^\nu
        + \tau_\mu P^\lambda_\rho \Omega_{\nu\lambda} v^\nu
        + P^\lambda_\rho \Omega_{\mu\lambda}
        + Q_{\nu\mu\rho} v^\nu \nonumber\\
      &= T_{\rho\mu\nu} v^\nu
        - T_{\mu\nu\rho} v^\nu
        + \tau_\mu \Omega_{\nu\rho} v^\nu
        + \Omega_{\mu\rho}
        - v^\lambda \tau_\rho \Omega_{\mu\lambda}
        + Q_{\nu\mu\rho} v^\nu \nonumber\\
      &= 2 T_{(\rho\mu)\nu} v^\nu
        - \tau_{(\mu} \Omega_{\rho)\nu} v^\nu
        + \Omega_{\mu\rho}
        + Q_{\nu\mu\rho} v^\nu \; ,
    \end{align}
    where we have used $h_{\kappa\nu} v^\nu = 0$ as well as
    antisymmetry of $\Omega$ and of $T$ in its last two indices.
    Antisymmetrising in $\rho$ and $\mu$ gives $2 S_{[\rho\mu]\nu}
    v^\nu = \Omega_{[\mu\rho]} = \Omega_{\mu\rho}$, which together
    with \eqref{eq:NC_form_via_diff} shows $\tilde{\Omega} = \Omega$.

    This finishes the proof of the classification theorem.  \qedhere
  \end{enumerate}
\end{proof}

\section{The principal bundle perspective}
\label{sec:bundle_persp}

As is well-known, connections on a Galilei manifold $(M,\tau,h)$ that
are compatible with $\tau$ and $h$ may equivalently be understood as
principal connections on a bundle of adapted frames
\cite{Kuenzle:1972,Schwartz:2023}.  Here, we wish to analyse this
relationship for general affine connections.  While the description of
affine connections as principal connections on the general linear
frame bundle is of course standard material, our analysis in terms of
objects on the subbundle of frames adapted to $\tau$ and $h$ is novel.
This will provide a different angle on some parts of our
classification result from \cref{sec:class}.

\subsection{The metric-compatible case}

First, we recall the standard construction for compatible
connections \cite{Kuenzle:1972,Schwartz:2023}.

\begin{definition}
  A \emph{(local) Galilei frame} on a Galilei manifold $(M,\tau,h)$ is
  a local frame $(\e_A) = (\e_t, \e_a) = (v, \e_a)$, $a = 1, \dots,
  n$, of vector fields on $M$ satisfying
  \begin{equation} \label{eq:Galilei_frame}
    \tau(v) = 1 \; , \quad
    h^{\mu\nu} = \delta^{ab} \e_a^\mu \e_b^\nu \; .
  \end{equation}
  The dual frame of one-forms then takes the form $(\e^A) = (\e^t,
  \e^a) = (\tau, \e^a)$.
\end{definition}

By direct calculation, one can check that the (pointwise) basis change
matrices between Galilei frames are precisely the elements of the
(orthochronous) homogeneous Galilei group $\Gal = \mathsf{O}(n)
\ltimes \R^n$, understood as a subgroup of $\GL(n+1)$ according to
\begin{equation} \label{eq:Gal_group_matrices}
  \Gal \ni (R,k) \mapsto
  \begin{pmatrix}
    1 & 0 \\
    k & R
  \end{pmatrix}
  \in \GL(n+1) \; .
\end{equation}
This shows the following:
\begin{proposition}
  Let $(M,\tau,h)$ be a Galilei manifold.  Of the general linear frame
  bundle $F(M) \to M$, we consider the subbundle $G(M)$ whose elements
  are bases $(v|_p, \e_a|_p)$ of tangent spaces $T_pM$ satisfying the
  pointwise equivalent of \eqref{eq:Galilei_frame}.  This $G(M)$ is a
  reduction of the structure group of $F(M)$ from $\GL(n+1)$ to
  $\Gal$; we call it the \emph{Galilei frame bundle} of $(M,\tau,h)$.
  By construction, the Galilei frames on $(M,\tau,h)$ are precisely
  the local sections of $G(M)$.  \qed
\end{proposition}

Since $G(M)$ is a reduction of the linear frame bundle, the tangent
bundle $TM$ is canonically isomorphic to the associated vector bundle
\begin{equation}
  E = G(M) \times_\Gal \R^{n+1} \; .
\end{equation}
A principal connection on $G(M)$, given by a connection form $\bomega
\in \Gamma(T^*G(M) \otimes \gal)$ valued in the Galilei algebra $\gal
= \mathrm{Lie}(\Gal) = \so(n) \oright \R^n$, then induces a covariant
derivative operator on $E$, i.e.\ (via the canonical isomorphism) a
covariant derivative operator $\nabla$ on $TM$---that is, an affine
connection on $M$.  This affine connection is then compatible with
$\tau$ and $h$; conversely, any compatible affine connection is
induced by a (unique) principal connection on the Galilei frame
bundle.

The explicit local description of $\nabla$ is as follows:
understanding a Galilei frame $(\e_A)$ on some open set $U \subset M$
as a local section $(\e_A) = \sigma$ of $G(M)$, we obtain the local
connection form $\omega = \sigma^*\bomega \in \Gamma(U, T^*M \otimes
\gal)$.  The Lie group inclusion \eqref{eq:Gal_group_matrices} induces
a Lie algebra inclusion
\begin{equation} \label{eq:gal_alg_matrices}
  \gal = \so(n) \oright \R^n \ni (X,k) \mapsto
  \begin{pmatrix}
    0 & 0 \\
    k & X
  \end{pmatrix}
  \in \gl(n+1) \; ,
\end{equation}
allowing us to view the Galilei algebra as a subalgebra of the Lie
algebra $\gl(n+1)$ of real-valued $(n+1)\times(n+1)$ matrices.  Hence
the local connection form may be viewed as a matrix-valued one-form:
decomposing it into its rotational and boost part $\omega =
(\tensor{\omega}{^a_b}, \varpi^a)$, with $(\tensor{\omega}{^a_b})$
taking values in $\so(n)$ and $(\varpi^a)$ in $\R^n$, it corresponds
to the matrix-valued form
\begin{equation}
  (\tensor{\omega}{^A_B})
  =
  \begin{pmatrix}
    0 & 0 \\
    \varpi^a & \tensor{\omega}{^a_b}
  \end{pmatrix}
  \in \Gamma(U, T^*M \otimes \gl(n+1)) \; .
\end{equation}
The affine connection $\nabla$ is then given by its action on the
frame fields $\e_A$ according to
\begin{subequations} \label{eq:local_conn_form}
\begin{align}
  \label{eq:local_conn_form_frame}
  \nabla \e_B &= \tensor{\omega}{^A_B} \otimes \e_A \; , \\
  \intertext{or on the dual frame fields $\e^A$ according to}
  \label{eq:local_conn_form_dual}
  \nabla \e^A &= - \tensor{\omega}{^A_B} \otimes \e^B \; .
\end{align}
\end{subequations}
Concretely, this reads
\begin{equation}
  \nabla v = \varpi^a \otimes \e_a \; , \quad
  \nabla \e_b = \tensor{\omega}{^a_b} \otimes \e_a \; .
\end{equation}

\subsection{The general case}
\label{sec:bundle_persp_general}

So far for the standard considerations for Galilei (i.e.\ compatible)
connections.  Now we want to generalise this to the case of general
affine connections $\nabla$ on $M$.  These correspond to principal
connections on the general linear frame bundle $F(M)$, i.e.\ to
connection forms $\bhatomega \in \Gamma(T^*F(M) \otimes \gl(n+1))$.
Restricting such a connection form to $G(M)$ (i.e.\ taking the
pullback along the inclusion $G(M) \hookrightarrow F(M)$), we obtain a
form $\bomega \in \Gamma(T^*G(M) \otimes \gl(n+1))$---i.e.\ a
$\gl(n+1)$-valued one-form on the Galilei frame bundle.  Note that
while this is not a connection form (it takes values in the wrong Lie
algebra), it still has the following properties:
\begin{enumerate}[label=(\roman*)]
\item $\bomega$ is equivariant with respect to the representation
  $\rho \colon \Gal \to \Aut(\gl(n+1))$ that arises by restricting the
  adjoint representation $\Ad \colon \GL(n+1) \to \Aut(\gl(n+1))$ to
  $\Gal \subset \GL(n+1)$, i.e.\ it satisfies
  \begin{equation}
    \mathrm{R}_g^* \bomega = \rho_{g^{-1}} \circ \bomega
  \end{equation}
  for all $g \in \Gal$, where $\mathrm{R}_g \colon G(M) \to G(M)$
  denotes the right translation by $g$.

\item Evaluating $\bomega$ on the fundamental vector field
  $\widetilde{(X,k)} \in \Gamma(TG(M))$ corresponding to an element
  $(X,k) \in \gal$, we obtain the matrix corresponding to $(X,k)$ by
  \eqref{eq:gal_alg_matrices}.
\end{enumerate}
Locally, the correspondence between the affine connection $\nabla$ and
the principal connection form $\bhatomega$ is again given via
\eqref{eq:local_conn_form}, where $(\e_A) = \hat\sigma \in \Gamma(U,
F(M))$ is a local frame of vector fields on $U \subset M$ and $\omega
= \hat\sigma^*\bhatomega \in \Gamma(U, T^*M \otimes \gl(n+1))$ is the
local connection form.  We now specialise to the local frame $(\e_A)$
being a Galilei frame, such that it may be viewed as a section
$\sigma$ of $G(M)$ and we can obtain the local connection form as
$\omega = \sigma^*\bomega$.

Since $(\e_A)$ is a Galilei frame, we have $\tau = \e^t$ and $h =
\delta^{ab} \e_a \otimes \e_b$, i.e.\ the frame components of $\tau$
and $h$ are
\begin{equation} \label{eq:tau_h_frame_comp}
  \tau_A = \delta_A^t \; , \quad
  h^{AB} = \delta^{ab} \delta^A_a \delta^B_b \; .
\end{equation}
Thus, the non-metricities $\hat Q = \nabla\tau$ and $Q = \nabla h$ of
$\nabla$---viewed as $T^*M$- and $(\bigvee^2TM)$-valued one-forms,
respectively---have frame components
\begin{equation} \label{eq:nabla_tau_frame_comp}
  \hat Q_A
  = (\nabla\tau)_A
  = \D\tau_A - \tensor{\omega}{^B_A} \tau_B
  = - \tensor{\omega}{^t_A}
\end{equation}
and
\begin{subequations} \label{eq:nabla_h_frame_comp}
\begin{align}
  Q^{AB}
  &= (\nabla h)^{AB} \nonumber\\
  &= \D h^{AB} + \tensor{\omega}{^A_C} h^{CB}
    + \tensor{\omega}{^B_C} h^{AC} \nonumber\\
  &= \tensor{\omega}{^A_c} \delta^{cb} \delta^B_b
    + \tensor{\omega}{^B_c} \delta^{ac} \delta^A_a \nonumber\\
  &= 2 \tensor{\omega}{^{(A}_c} \delta^{B)}_b \delta^{bc} \; .
\end{align}
Separating the latter into timelike and spacelike components, we have
\begin{equation} \label{eq:nabla_h_frame_comp_expl}
  Q^{tt} = 0 \; , \quad
  Q^{ta} = Q^{at} = \omega^{ta} \; , \quad
  Q^{ab} = 2 \omega^{(ab)} \; ,
\end{equation}
\end{subequations}
where we have raised spatial frame indices with $\delta^{ab}$, i.e.\
$\omega^{Ab} = \delta^{bc} \tensor{\omega}{^A_c}$.  Note that by this
form of the non-metricities, we have re-obtained the first of the
identities \eqref{eq:conn_class_ids}, relating the timelike part of
$Q$ to the spacelike part of $\hat Q$: in terms of frame components,
this identity reads $\tau_A Q^{AB} = -\hat Q_A h^{AB}$, which by
inserting \eqref{eq:tau_h_frame_comp} becomes $Q^{tB} = -\hat Q_a
\delta^{ab} \delta^B_b$.  This is satisfied by virtue of
\eqref{eq:nabla_tau_frame_comp} and
\eqref{eq:nabla_h_frame_comp_expl}, since $Q^{tt} = 0$ and $Q^{tb} =
\omega^{tb} = -\hat Q_a \delta^{ab}$.

The frame components of the torsion of $\nabla$ we can compute by
Cartan's first structure equation for affine connections in terms of
the dual frame as
\begin{equation}
  T^A = \D\e^A + \tensor{\omega}{^A_B} \wedge \e^B \; .  
\end{equation}
For the temporal torsion, this yields
\begin{subequations}
\begin{align}
  T^t &= \D\tau + \tensor{\omega}{^t_B} \wedge \e^B \; , \\
  \intertext{which with \eqref{eq:nabla_tau_frame_comp} becomes}
  T^t &= \D\tau - \hat Q_B \wedge \e^B \; .
\end{align}
\end{subequations}
This is precisely the translation of the second identity of
\eqref{eq:conn_class_ids} into the present language: the coordinate
components of $\hat Q_B \wedge \e^B$ are $(\hat Q_B \wedge
\e^B)_{\mu\nu} = 2 \hat Q_{[\mu|\rho}^{\vphantom{B}} \e_B^\rho
\e^B_{|\nu]} = 2 \hat Q^{\vphantom{\rho}}_{[\mu|\rho}
\delta^\rho_{|\nu]} = 2 \hat Q_{[\mu\nu]}$.

By direct calculation, one can show that exactly as in the case of
compatible connections, the Newton--Coriolis form $\Omega$ of $\nabla$
with respect to the unit timelike vector field $v = \e_t$ from the
Galilei frame is given by
\begin{equation} \label{eq:NC_form_frame}
  \Omega = \delta_{ab} \varpi^a \wedge \e^b \; ,
\end{equation}
where $\varpi^a = \tensor{\omega}{^a_t}$.

Decomposing the local connection form as
\begin{equation} \label{eq:loc_conn_form_decomp}
  (\tensor{\omega}{^A_B})
  =
  \begin{pmatrix}
    \tensor{\omega}{^t_t} & \tensor{\omega}{^t_b} \\
    \varpi^a & \tensor{\omega}{^a_b}
  \end{pmatrix}
  =
  \begin{pmatrix}
    0 & 0 \\
    \varpi^a & \omega^{[ac]} \delta_{bc}
  \end{pmatrix}
  +
  \begin{pmatrix}
    \tensor{\omega}{^t_t} & \tensor{\omega}{^t_b} \\
    0 & \omega^{(ac)} \delta_{bc}
  \end{pmatrix}
\end{equation}
into its part taking values in the Galilei algebra $\gal \subset
\gl(n+1)$ (understood as a subalgebra according to
\eqref{eq:gal_alg_matrices}) and the rest, this `rest' is precisely
what prevents the form from taking values in $\gal$, i.e.\ what
prevents $\nabla$ from being compatible with $\tau$ and $h$.  In fact,
the `rest' consists of $\tensor{\omega}{^t_B}$ and the symmetrised
part of the spatial connection form $\tensor{\omega}{^a_b}$, i.e.\
precisely the non-metricities $\hat Q = \nabla\tau$ and $Q = \nabla h$
according to \eqref{eq:nabla_tau_frame_comp},
\eqref{eq:nabla_h_frame_comp}.  This may also be understood from a
more global perspective on the Galilei frame bundle $G(M)$, as
follows.

Recall that by $\rho \colon \Gal \to \Aut(\gl(n+1))$ we denote the
restriction of the adjoint representation of $\GL(n+1)$ to $\Gal$.  By
construction, the Galilei algebra $\gal \subset \gl(n+1)$ is an
invariant subspace of the representation $\rho$ (since on $\gal$,
$\rho$ is simply the adjoint representation of $\Gal$).  Hence, we
obtain a quotient representation
\begin{equation}
  \tilde{\rho} \colon \Gal \to \GL(\gl(n+1)/\gal)
\end{equation}
of $\Gal$ on the quotient vector space $\gl(n+1)/\gal$ (note that due
to $\gal \subset \gl(n+1)$ not being an ideal, this quotient is not a
Lie algebra in any natural way).  Therefore, from the form $\bomega
\in \Gamma(T^*G(M) \otimes \gl(n+1))$ on the Galilei frame bundle
encoding the affine connection, by composition with the natural
projection $\gl(n+1) \to \gl(n+1)/\gal$ we obtain a form that is
tensorial with respect to the quotient representation,
\begin{equation}
  \bomega + \gal \in \Omega^1_{\tilde{\rho}} (G(M),\gl(n+1)/\gal) \;.
\end{equation}
(Note that this form indeed vanishes on vertical vectors, since on
vertical vectors $\bomega$ takes values in $\gal$.)  Pulled back to
$M$ by a local section, this $\tilde{\rho}$-tensorial form corresponds
precisely to the second part of the local connection form in the
decomposition \eqref{eq:loc_conn_form_decomp}.  Hence, it encodes the
non-metricities $\hat Q$ and $Q$ in a global fashion on $G(M)$.

Summarising all of the above considerations, we have the following:
\begin{theorem} \label{thm:bundle_persp_general}
  Let $(M,\tau,h)$ be a Galilei manifold, $G(M)$ its Galilei frame
  bundle and $F(M)$ the general linear frame bundle of $M$.  An affine
  connection $\nabla$ on $M$ corresponds to a connection form
  $\bhatomega$ on $F(M)$, or equivalently---taking the pullback of
  $\bhatomega$ along the inclusion $G(M) \hookrightarrow F(M)$---to a
  form $\bomega \in \Gamma(T^*G(M) \otimes \gl(n+1))$ satisfying the
  following:
  \begin{enumerate}[label=(\roman*)]
  \item $\bomega$ is equivariant with respect to the representation
    $\rho \colon \Gal \to \Aut(\gl(n+1))$ that arises by restricting
    the adjoint representation of $\GL(n+1)$ to $\Gal$.
  \item Evaluated on the fundamental vector field $\widetilde{(X,k)}
    \in \Gamma(TG(M))$ of an element $(X,k) \in \gal$, the form
    $\bomega$ yields the matrix corresponding to $(X,k)$ in $\gl(n+1)$
    by \eqref{eq:gal_alg_matrices}.
  \end{enumerate}
  Composing $\bomega$ with the projection $\gl(n+1) \to
  \gl(n+1)/\gal$, we obtain a form $\bomega + \gal$ on $G(M)$ with
  values in $\gl(n+1)/\gal$ that is tensorial with respect to the
  quotient representation $\Gal \to \GL(\gl(n+1)/\gal)$.  This encodes
  the non-metricities $\nabla\tau$ and $\nabla h$ of the affine
  connection $\nabla$.

  Pulling back the form $\bomega$ along a local Galilei frame $(\e_A)
  = (v, \e_a)$ on $U \subset M$, we obtain the local connection form
  $\omega = (\tensor{\omega}{^A_B}) \in \Gamma(U, T^*M \otimes
  \gl(n+1))$.  In terms of this, the non-metricities $\hat Q =
  \nabla\tau$ and $Q = \nabla h$ are given by
  \eqref{eq:nabla_tau_frame_comp}, \eqref{eq:nabla_h_frame_comp}, and
  the Newton--Coriolis form of $\nabla$ with respect to $v$ is given
  by \eqref{eq:NC_form_frame}.  \qed
\end{theorem}

Finally, as in \cref{sec:class}, we are going to briefly discuss the
curvature of the connection.  As for any principal connection, the
curvature form of $\bhatomega$ on the general linear frame bundle is a
two-form $\boldsymbol{\hat R} \in \Gamma(\bigwedge^2 T^*F(M) \otimes
\gl(n+1))$ given by the structure equation
\begin{subequations}
\begin{align}
  \boldsymbol{\hat R}
  &= \D\bhatomega + \frac{1}{2} [\bhatomega \wedge \bhatomega] \; . \\
  \intertext{Pulling this back along a local frame $(\e_A) = \hat\sigma \in
  \Gamma(U, F(M))$, we obtain the local curvature form $R = \hat\sigma^*
  \boldsymbol{\hat R} \in \Gamma(U, T^*M \otimes \gl(n+1))$.  Due to
  $\gl(n+1)$ being a matrix Lie algebra, the structure equation for the
  curvature form may be written in matrix component form as}
  \tensor{R}{^A_B}
  &= \D\tensor{\omega}{^A_B}
    + \tensor{\omega}{^A_C} \wedge \tensor{\omega}{^C_B} \; ,
\end{align}
which is Cartan's second structure equation for affine connections.
Explicitly decomposing into timelike and spacelike entries, we have
\begin{align} \label{eq:curv_form_expl}
  (\tensor{R}{^A_B})
  &= (\D\tensor{\omega}{^A_B}
    + \tensor{\omega}{^A_C} \wedge \tensor{\omega}{^C_B}) \nonumber\\
  &=
  \begin{pmatrix}
    \D\tensor{\omega}{^t_t} + \tensor{\omega}{^t_c} \wedge \varpi^c
    & \D\tensor{\omega}{^t_b}
    + \tensor{\omega}{^t_t} \wedge \tensor{\omega}{^t_b}
    + \tensor{\omega}{^t_c} \wedge \tensor{\omega}{^c_b}\\
    \D\varpi^a + \varpi^a \wedge \tensor{\omega}{^t_t}
    + \tensor{\omega}{^a_c} \wedge \varpi^c 
    & \D\tensor{\omega}{^a_b} + \varpi^a \wedge \tensor{\omega}{^t_b}
    + \tensor{\omega}{^a_c} \wedge \tensor{\omega}{^c_b}
  \end{pmatrix} \; .
\end{align}
\end{subequations}
The curvature further satisfies the Bianchi identities, which are
valid for any connection on any reduction of the general linear frame
bundle, and in form language read
\begin{subequations} \label{eq:Bianchi_form}
\begin{align}
  \tensor{R}{^A_B} \wedge \e^B
  &= (\D^\omega T)^A
    = \D T^A + \tensor{\omega}{^A_B} \wedge T^B \; , \\
  \D \tensor{R}{^A_B} + \tensor{\omega}{^A_C} \wedge \tensor{R}{^C_B}
  &= \tensor{(\D^\omega R)}{^A_B}
    = 0 \; .
\end{align}
\end{subequations}
As we did in \cref{sec:class} for their coordinate components, we
leave a more explicit spelling out of the curvature form
\eqref{eq:curv_form_expl} and the Bianchi identities
\eqref{eq:Bianchi_form} in terms of the independent fields in terms of
which the connection may be classified to future work.

\subsection{Formal Newtonian limits}

The principal bundle perspective is also enlightening regarding the
(formal) Newtonian limit from Lorentzian to Galilei geometry.  As is
well-known \cite{Kuenzle:1976,Ehlers:1981a, Ehlers:1981b}, starting
with a family of Lorentzian metrics depending on a parameter $c$
representing the speed of light, the degenerate limit $c\to\infty$
yields the metric structures $\tau$ and $h$ of a Galilei manifold
(under specific assumptions on the $c$ dependence, see, e.g., ref.\
\cite{Kuenzle:1976} for details).  In this limit, the orthonormal
frame bundle of the Lorentzian manifold goes over to the Galilei frame
bundle, realising the İnönü--Wigner contraction
\cite{Inonu.Wigner:1953} from the Lorentz to the homogeneous Galilei
algebra on the level of structure groups.  Concretely, one may also
show that under quite weak assumptions Lorentzian orthonormal frames
converge pointwise to Galilei frames \cite{Schwartz.vonBlanck:2024}.

Concerning compatible affine connections, any Lorentzian
metric-compatible connection which has a regular $c\to\infty$ limit
will converge to a Galilei connection, i.e.\ a connection compatible
with $\tau$ and $h$.  This is most easily shown using local connection
forms with respect to Lorentzian orthonormal frames and Galilei
frames, respectively \cite{Schwartz:2023}.  Further, using the
Lie-algebraic perspective, the Newtonian limit may be understood in an
even more explicitly gauge-theoretic form: the geometric objects of
(standard) post-Newton--Cartan geometry, arising from formal power
series expansion of Lorentzian geometric objects in $c^{-1}$, may
alternatively be obtained by applying a `gauging procedure' to a
specific Lie algebra---namely one that arises as a so-called Lie
algebra expansion of the Poincaré algebra in the parameter $c^{-1}$
\cite{Hansen.EtAl:2020}.

The Galilei frame bundle point of view on general affine connections
on Galilei manifolds, as developed in \cref{sec:bundle_persp_general}
and formulated in particular in \cref{thm:bundle_persp_general}, can
be expected to allow for a similar understanding of the Newtonian
limit of connections in general metric-affine Lorentzian geometry, and
corresponding theories of gravity.  Extending beyond just the
connection, in the end one would expect that general metric-affine
post-Newton--Cartan geometry may be obtained as a `gauging' of a
$c^{-1}$ expansion of a suitable `kinematical split' of the affine
group into a Poincaré and a non-metric part.

\section{Conclusion}
\label{sec:conclusion}

In this paper, we have developed the classification of affine
connections on Galilei manifolds in terms of their torsion,
Newton--Coriolis form with respect to a fixed unit timelike vector
field, and non-metricities.  We explained the interpretation of our
general classification result as giving an affine isomorphism between
the space of affine connections and a specific vector space of tensor
fields, thereby making precise the possibility of specifying a
connection in terms of tensor fields that may be freely prescribed.
This offers an important generalisation of the well-understood case of
Galilei connections, i.e.\ metric-compatible ones
\cite{Bekaert.Morand:2016,Geracie.EtAl:2015,Bergshoeff.EtAl:2023}.  We
also discussed this problem in the language of local connection forms
with respect to Galilei frames, relating this to a globalised
principal bundle point of view.

Our result enables the systematic study of Newton--Cartan-like limits
of general metric-affine theories of `relativistic' gravity,
generalising previous work along these lines concerning the (metric)
teleparallel and symmetric teleparallel equivalents of \acronym{GR}
\cite{Read.Teh:2018,Schwartz:2023,Wolf.EtAl:2024}.  Further, it will
allow for an extension of such studies beyond consideration of the
limit, namely the development of a coordinate-free geometric
formulation of the post-Newtonian expansion of such theories---an
essential part of a comprehensive understanding of post-Newtonian
gravity.

\section*{Acknowledgements}

I wish to thank Arian von Blanckenburg, Domenico Giulini, James Read,
Miguel Sánchez, Quentin Vigneron, and William Wolf for valuable
discussions.

\printbibliography

\appendix

\section{The special Galilei connection defined by a vector field}
\label{sec:app_spec_conn}

To make our proof of the classification theorem fully self-contained,
in the following we give a proof of \cref{lem:nablav}.
\begin{proof}[Proof of \cref{lem:nablav}]
  Let $(M,\tau,h)$ be a Galilei manifold and $v$ a unit timelike
  vector field on it, and consider the connection $\nablav$ with
  coefficients
  \begin{equation}
    \Gammav^\rho_{\mu\nu}
    = v^\rho \partial_\mu \tau_\nu
    + \frac{1}{2} h^{\rho\sigma} (\partial_\mu h_{\nu\sigma}
      + \partial_\nu h_{\mu\sigma} - \partial_\sigma h_{\mu\nu}) \; .
  \end{equation}
  We want to show that $\nablav\tau$, $\nablav h$, and the
  Newton--Coriolis form of $\nablav$ with respect to $v$ vanish.

  First, using $\tau_\rho v^\rho = 1$ and $\tau_\rho h^{\rho\sigma} =
  0$ we directly obtain
  \begin{equation}
    \nablav_\mu \tau_\nu 
    = \partial_\mu \tau_\nu    
      - \Gammav^\rho_{\mu\nu} \tau_\rho 
    = \partial_\mu \tau_\nu - \partial_\mu \tau_\nu = 0 \; .
  \end{equation}

  Next, using the Leibniz rule we compute
  \begin{align}
    \nablav_\mu h^{\rho\sigma}
    &= \partial_\mu h^{\rho\sigma}
      + \Gammav^\rho_{\mu\lambda} h^{\lambda\sigma}
      + \Gammav^\sigma_{\mu\lambda} h^{\rho\lambda} \nonumber\\
    &= \partial_\mu h^{\rho\sigma}
      + 2 \Gammav^{(\rho}_{\mu\lambda}
        h^{\sigma)\lambda}_{\vphantom{\lambda}} \nonumber\\
    &= \partial_\mu h^{\rho\sigma}
      + 2 \underbrace{(\partial_\mu \tau_\lambda)
          h^{\lambda(\sigma}}_{= - \tau_\lambda \partial_\mu
          h^{\lambda(\sigma}}
        v^{\rho)}
      + h^{\lambda(\sigma} h^{\rho)\kappa} (\partial_\mu h_{\lambda\kappa}
        +{} \underbrace{\partial_\lambda h_{\mu\kappa}
          - \partial_\kappa h_{\mu\lambda}}_{= \cancelto{0}{2
          \partial_{[\lambda} h_{\kappa]\mu}}}) \nonumber\\
    &= \partial_\mu h^{\rho\sigma}
      - 2 \underbrace{\tau_\lambda v^{(\rho}}_{\mathclap{=
          \delta^{(\rho}_\lambda - P^{(\rho}_\lambda}}
        \partial_\mu h^{\sigma)\lambda}
      + h^{\lambda\sigma} \underbrace{h^{\rho\kappa} \partial_\mu
        h_{\lambda\kappa}}_{\mathclap{= \partial_\mu P^\rho_\lambda
        - h_{\lambda\kappa} \partial_\mu h^{\rho\kappa}}} \nonumber\\
    &= -\cancel{\partial_\mu h^{\rho\sigma}}
      + \bcancel{2 P^{(\rho}_\lambda \partial_\mu
        h^{\sigma)\lambda}_{\vphantom{\lambda}}}
      + \underbrace{h^{\lambda\sigma} \partial_\mu
        P^\rho_\lambda}_{\mathclap{= 
        \cancel{\partial_\mu h^{\rho\sigma}} 
        - \bcancel{P^\rho_\lambda \partial_\mu h^{\lambda\sigma}}}}{}
      - \bcancel{P^\sigma_\kappa \partial_\mu h^{\rho\kappa}}
      \nonumber\\
    &= 0 \; .
  \end{align}

  Finally, the components of the Newton--Coriolis form of $\nablav$
  with respect to $v$ are
  \begin{align}
    \Omega_{\mu\nu}
    &= 2 h_{\rho[\nu} \nablav_{\mu]} v^\rho \nonumber\\
    &= 2 h_{\rho[\nu} \partial_{\mu]} v^\rho
      + 2 h_{\rho[\nu} \Gammav^\rho_{\smash{\mu]\kappa}} v^\kappa
      \nonumber\\
    &= -2 v^\rho \partial_{[\mu} h_{\nu]\rho}
      + v^\kappa P^\sigma_{\smash{[\nu}}
        (\underbrace{\partial_{\mu]} h_{\kappa\sigma}}
          _\text{Leibniz}
        {}+ \underbrace{\partial_{|\kappa|} h_{\mu]\sigma}}
          _\text{Leibniz}
        {}- \underbrace{\partial_{|\sigma|} h_{\mu]\kappa}}
          _\text{defn.\ $P^\sigma_\nu$})
      \nonumber\\
    &= -\cancel{2 v^\rho \partial_{[\mu} h_{\nu]\rho}}
      + v^\kappa (\cancel{\partial_{[\mu} h_{\nu]\kappa}}
        - \cancelto{0}{h_{\kappa\sigma}} \partial_{[\mu}
          P^\sigma_{\smash{\nu]}}
        - h_{\sigma[\mu} \smash{\underbrace{\partial_{|\kappa|}
          P^\sigma_{\smash{\nu]}}}
          _{\mathclap{= - \tau_{\nu]} \partial_\kappa v^\sigma
          - \cancelto{0}{v^\sigma \partial_{|\kappa|} \tau_{\nu]}}}}}
      \nonumber\\
    &\qquad - \cancel{\partial_{[\nu} h_{\mu]\kappa}}
        + v^\sigma \tau_{[\nu} \partial_{|\sigma|} h_{\mu]\kappa})
      \nonumber\\
    &= v^\kappa h_{\sigma[\mu} \tau_{\nu]} \partial_\kappa v^\sigma
      + v^\sigma \tau_{[\nu} \underbrace{v^\kappa \partial_{|\sigma|}
        h_{\mu]\kappa}}_{= - h_{\mu]\kappa} \partial_\sigma v^\kappa}
      \nonumber\\
    &= 0 \; .
  \end{align}
\end{proof}

\end{document}
